\documentclass[preprint]{aastex}

\usepackage{graphicx}
\usepackage{amsmath}
\usepackage{amssymb}
\usepackage{mathrsfs}

\newcommand{\Tfl}{$\theta_{fl}$}
\newcommand{\EQ}[1]{Eq.(\ref{EQ #1})}  

\begin{document}
\title{Polarization \& light curve variability: the ``patchy-shell'' model}
\author{Ehud Nakar and Yonatan Oren}
\affil{
    The Racah Institute of Physics, The Hebrew University,
    Jerusalem, Israel, 91904}

\begin{abstract}
Recent advances in early detection and detailed monitoring of
gamma-ray burst (GRB)
afterglows have revealed variability in some afterglow light
curves. One of the leading models for this behavior is the patchy
shell model. This model attributes the variability to random
angular fluctuations in the relativistic jet energy. These
non axisymmetric fluctuations should also impose variations in the
degree and angle of polarization that are correlated to the light
curve variability. In this letter we present a solution of the
light curve and polarization resulting from a given spectrum of
energy fluctuations. We compare light curves produced using this
solution with the variable light curve of GRB 021004, and we show that
the main features in both the light curve and the polarization
fluctuations are very well reproduced by this model. We use our
results to draw constraints on the characteristics of the energy
fluctuations that might have been present in GRB 021004.
\end{abstract}

\section{Introduction}
Within the Fireball model for gamma-ray bursts (GRB
s) (Piran 2000, M\'esz\'aros 2002), the emission process in the
optical and X-ray bands during the afterglow (AG) is most likely an
optically thin, slow cooling synchrotron. 
Under the simplifying assumptions of spherical or scale free axial
symmetry, this model predicts a smooth,
broken power-law light curve. Until recently most all of the observed
AGs exhibited a light curve conforming to the above predictions of the
model. However, 
recently several observed AGs (mainly GRB 021004 and GRB 030329)
showed variable light curves that can be interpreted as
fluctuations superimposed on a power law decay. These two AGs were
recorded with especially good resolution
and accuracy, and they were detected very shortly after the GRB. Thus
it is not clear to what extent the compatibility of earlier AG
observations with a broken power-law indicates an intrinsic
agreement (as opposed to sparse sampling).

Fluctuating light curves were predicted by various
models. The most plausible models suggest variations in the
blast wave's energy or in the external density. These variations
can be (locally) spherically symmetric as in the energy
fluctuated refreshed shocks model (Rees \& M\'esz\'aros 1998, Kumar \&
Piran 2000a, Sari \& M\'esz\'aros 2000), or they can be aspherical 
variations of the energy (as in the patchy-shell model; Kumar \&
Piran 2000b) or the external density  (Wang \& Loeb 2000, Lazzati et
al. 2002, Nakar et al. 2003). Motivated by
the clear evidence for deviations from axisymmetry in at least one
burst (GRB 021004) we
focus our attention on the aspherical models. In this letter, we
investigate the patchy-shell model.

In the patchy-shell model the energy per solid angle of the blast
wave displays angular variations. These energy variations induce
fluctuations in the AG
light curve. Because of the non axisymmetric nature of the
energy variations they also impose variations in the degree and
angle of polarization that are correlated to the light curve
variability (Granot \& K\"onigel 2003). We calculate the light
curve and the polarization resulting from a given spectrum of energy
fluctuations. We show that generally the variability time scale
$\Delta T$
behaves as $\Delta T \sim T$, and the amplitude envelope decays as
$T^{-3/8}$, where $T$ is the time in the observer frame. We also find
a correlation and time delay 
between light-curve variations in different spectral bands.
Current observations restrict the amplitude of energy fluctuations
to be less than a factor of $10$ (otherwise we would not
expect the observed narrow distribution of $\gamma$-ray emission
energy; Frail et al. 2001). We show here that such energy variations
are consistent with the observations, namely that they can produce both variable and smooth
light curves, depending on the observer location. Piran (2001)
even argues that such fluctuations may solve the puzzle of why the
energy emitted in $\gamma$-rays seems larger than the kinetic
energy that remains in the blast-wave, whereas the opposite is expected.

GRB 021004 has all the properties expected from a non spherically
symmetric burst: Its AG displays steep decays on time scales that
cannot be obtained in a 
spherically symmetric model (Nakar \& Piran 2003) and  its
polarization shows rapid fluctuations in the polarization
angle and degree (Rol et al. 2003). These fluctuations cannot be
explained by any of the current models, providing further
indication that the radiation source is non axisymmetric
(Lazzati et al. 2003). These fluctuations in the polarization were
even predicted by Granot \& K\"onigel (2003) (based on the variable
light curve and the expected axisymmetry break) prior to the
observational report. We demonstrate 
that the patchy-shell model is capable of explaining the
light-curve and polarization (amplitude and angle) of GRB 021004
and we determine the properties of the angular energy distribution
that can account for the observed behavior.

In \S2 we calculate the light curve and polarization from a patchy
shell. In \S3 we find
an energy profile that reproduces the observed light curve and
polarization of GRB 021004. We draw our conclusions in \S4.

\section{The light curve and polarization calculation}

We calculate the observed light curve, degree of polarization and
polarization angle, resulting from a synchrotron emission of an
adiabatic blast wave with angular fluctuations in the energy,
$E=E(R,\theta ,\phi)$, where $E$ is the 
energy per solid angle \footnote{Throughout the paper we use
spherical coordinates with the origin at the center of the blast,
$\theta$ is the polar angle w.r.t the line of sight and $\phi$ is
the azimuthal angle}. We assume that the energy of both the
electrons and the magnetic field are in constant
equipartition with the total internal energy of the shocked fluid
and we take the circum-burst medium density as a constant
(interstellar medium).
Based on the thin shell nature of the Blandford \& Mckee (1976)
solution ($R/\Delta R\approx 16{\gamma}^2$, where $\gamma$ is the
Lorentz factor of the freshly shocked fluid), we approximate the
radiating region to be only the instantaneous shock front.

In interstellar medium (ISM), an adiabatic blast wave propagates at a
Lorentz factor $\Gamma \propto 
R^{-3/2}$ ($\Gamma=\sqrt{2}\gamma$). Since $\theta_s$,
the angular size of regions at radius $R$ causally connected by
sound waves that propagate at 
$\beta_s=1/\sqrt{3}$ (in the fluid rest
frame) grows as $d\theta_s=\beta_s dR/(\Gamma R)$, we obtain:
\begin{equation}
\theta_s(R)=\frac{2\beta_s}{3}\frac{1}{\Gamma}\approx\frac{1}{4\gamma}.
\label{EQ theta s}
\end{equation}
As long as the typical angular size
of the energy fluctuations, $\theta_{fl}$, is larger than
$\theta_s$, the energy profile  is ``frozen'' in time ($E=E(\theta
,\phi)$). Moreover, Kumar and Granot (2003) have shown that the
actual transversal velocities of the fluid may be much smaller than
the speed of sound. Consequently, the ``frozen shell''
approximation may remain valid even when $\theta_s \gtrsim
\theta_{fl}$. Following these arguments, we carry out our
calculation using the ``frozen shell'' approximation, which
facilitates our calculation considerably, as we can treat each
element of solid angle as part of a homogeneous sphere.

Under the above approximations, the contribution to the flux
per unit of observer frequency $\nu$ from an element of
solid angle $d\Omega$ at radius $R$ is given by
(Sari 1998):
\begin{equation}
dF_{\nu}(R,\theta,\phi) \propto L'_{\nu \gamma(1-\beta \cos
\theta)} [\gamma(1-\beta Cos\theta)]^{-3}d\Omega, \label{EQ dFnu1}
\end{equation}
where $L'_{\nu'}(R)$ is the luminosity of the solid angle element
in the fluid rest frame. Calculating $L'_{\nu'}$ following the
procedure of Sari, Piran and Narayan (1998) for the slow cooling
regime we obtain:
\begin{equation}
dF_{\nu}(T,\theta, \phi)\propto \left( \frac{1+y}{\gamma }\right)
^{-(3+\alpha)} d\Omega \left\{ \begin{array}{c}
E^{\frac{3}{4}}T^{\frac{3}{4}}(y+1/8)^{-\frac{3}{4}}\; \ \ \ \ \ \ \ \ \ \ \ \ \nu <\nu _{m}\\
E^{\frac{11+3p}{16}}T^{\frac{15-9p}{16}}(y+1/8)^{\frac{9p-15}{16}}\;
\ \ \ \ \ \ \ \nu _{m}<\nu <\nu _{c}\\
E^{\frac{6+3p}{16}}T^{\frac{14-9p}{16}}(y+1/8)^{\frac{9p-14}{16}}\;
\ \ \nu _{c}<\nu
\end{array}\right.,
\label{EQ dFnu2}
\end{equation}
where $\alpha$, the spectral power law index, is equal to
$-1/3,(p-1)/2$ and $p/2$ in each segment respectively. We have
used here the definition $y\equiv (\gamma\theta)^2$ and $\theta
\ll 1$. We also use the adibacity of the blast wave ($E \propto
R^3 \gamma^2$) that yields (Sari 1998) $R \propto
(ET/(y+1/8))^{1/4}$ for any solid angle element. The angular
dependence is implicit in $E$ and $y$ through the expression:
\begin{equation}
\gamma =3.6(1+8y)^{3/8} (E/10^{52})^{1/8} n^{-1/8} T_d^{-3/8}.
\label{EQ gamma}
\end{equation}
Now, the total observed flux, $F(T)$, is easily calculated by
integration over the solid angle.

Having obtained the flux contribution per solid angle element we
are able to calculate the linear polarization ($V=0$) as well. The
total stokes parameters are simply the average of the local stokes
parameters weighted by the flux \footnote{In the AG the relevant
polarization is instantaneous, thus it is  weighted by the flux,
see Nakar, Piran \& Waxman (2003) and Granot (2003).}:
\begin{equation}
\frac{\left\{ \begin{array}{c}Q \\U \\
\end{array}\right\}}{I\Pi _{synch}}=
\frac{\int dF_{\nu}\Pi(y)\left\{ \begin{array}{c} \cos(2\theta
_{p}) \\ \sin(2\theta _{p})\\
\end{array}\right\} }{\int dF_{\nu}} ,
\label{EQ Q U}
\end{equation}
where $\Pi_{synch}$ is the polarization of synchrotron
emission in the fluid frame at the relevant power-law segment
(Granot 2003, for $\nu_m<\nu<\nu_c$ it is $(p+1)/(p+7/3)$),
$\theta_p$ is the polarization angle and $\Pi$ is the observed
local polarization relative to $\Pi_{synch}$. Note that the
integration over $dF$ is actually an integration over $d\Omega$.
$\Pi$ and $\theta_p$ at each element depend on two factors: (i)
The Lorentz boost of a photon emitted from that element and
reaching the observer, which  depends on $y$ (note that $y$ depends on
$T$ and $E$, and thus on 
$\theta$ \& $\phi$ (\EQ{gamma}), and (ii) the magnetic field
configuration - random or uniform. A Random $B$ is described by the
level of anisotropy, $b \equiv 2\langle B_\parallel^2 \rangle
/\langle B_\perp^2\rangle$ (Granot \& K\"onigel 2003), where
$B_\parallel$ is a random component in the plane of the shock and
$B_\perp$ is the component parallel to the propagation of
the fluid. In this case $\Pi(y)/\Pi_{synch} \approx
2y(b-1)/((1+y)^2+2y(b-1))$ (Gruzinov 1999, Sari 1999, Granot 2003)
and $\theta_p$ is radial [tangential] for $b<1$ [$b>1$]. In
Uniform $B$, $\Pi=1$ is constant and $\theta_p$ is given in Granot
\& K\"onigel (2003).

Although we are concerned with angular fluctuations, it is
illuminating to consider first the spherically symmetric case. In
this case the contribution to the observed flux at a given
observer time is concentrated within a ring centered on the line
of sight (Naturally all observed quantities here are independent
of $\phi$). The flux under these conditions is given by a self
similar function of $\theta$, $\xi(\theta) \equiv
dF_{\nu}/d\theta$, when $\theta$ is measured in units of
$(T^3nm_pc^5/E)^{1/8}$ and its height is normalized. Fig. (1)
depicts $\xi$ for the three different spectral power law segments.
$\xi$ is localized with FWHM of 0.5[1]$\theta_{max}$ for
$\nu>\nu_m[\nu<\nu_m]$, where $\theta_{max}$ is the angle of
maximum $\xi$. $\xi$ depends weakly (as $E^{1/8}$) on the energy,
so in the case of a non-spherical energy distribution, as long as
the energy variations are not large the shape of the observed ring
is only mildly distorted. This analysis enables us to derive a
constraint on the time scale of fluctuations in the light curve. A
significant fluctuation can occur only after the ring is
displaced such that it covers an essentially new region. As the
FWHM is of the order of $\theta_{max}$, and with the  power law
dependence of $\theta_{max}$ on $T$, this takes place on time
scales of the order of $T$. Thus, the time scale $\Delta T$ for fluctuations
in a light curve produced by a patchy shell obeys the
simple rule $\Delta T \gtrsim T$. One can understand this result
in terms of angular and radial times. Although the angular time of
a spot may be $< T$, its radial time, which is determined by the
time over which the ring crosses a spot, is of the order of $T$.
Naturally, no fluctuations are expected as long as
$\theta_{fl}>\theta_{max}$. Hence, $\Delta T \sim
max\{T,0.025[0.05](E/10^{52})^{1/3}n^{-1/3}(\theta_{fl}/0.03)^{8/3}\rm
days\}$  for $\nu>\nu_m[\nu<\nu_m]$. Also, in case \Tfl\ is much
smaller than $\theta_{max}$ (at late times, in case the frozen
shell approximation still holds) we would expect to see small time
scale fluctuations, which ``survive'' the smoothing effect,
superimposed on the main features. However these fluctuations turn
out to be so weak as to be completely hidden under the larger
scale structures.

A few other properties that can be drawn from the behavior of $\xi$
(see Fig. 1) are: (i) The value of the Lorentz factor at
$\theta_{max}$, $\gamma_{max}
\equiv \gamma(\theta_{max})$, which can be regarded as the
characteristic Lorentz factor at time $T$ for
$\nu>\nu_m[\nu<\nu_m]$ is:
\begin{equation}
\gamma_{max}=7.7[5] (E/10^{52})^{1/8}n^{-1/8}T_d^{-3/8}. \label{EQ
gamma max}
\end{equation}
(ii) The relation $\gamma_{max} \theta_{max} =0.9 [0.45]$ for
$\nu>\nu_m[\nu<\nu_m]$ is constant, owing to the self-similarity of
$\xi$. (iii) The overall amplitude of the fluctuations decreases
as the square root of the number of observed spots and is $\propto
\theta_{fl} /\theta_{max} \propto T^{-3/8}$ (Nakar et al. 2003).
(iv) The fluctuations at the three power law segments
$\nu_m<\nu<\nu_c$, $\nu_c<\nu$ and $\nu<\nu_m$ are correlated, but
the first two are simultaneous while the fluctuations in the third
segment are delayed relative to them by approximately $\Delta T$.

\begin{figure}[h]
\plotone{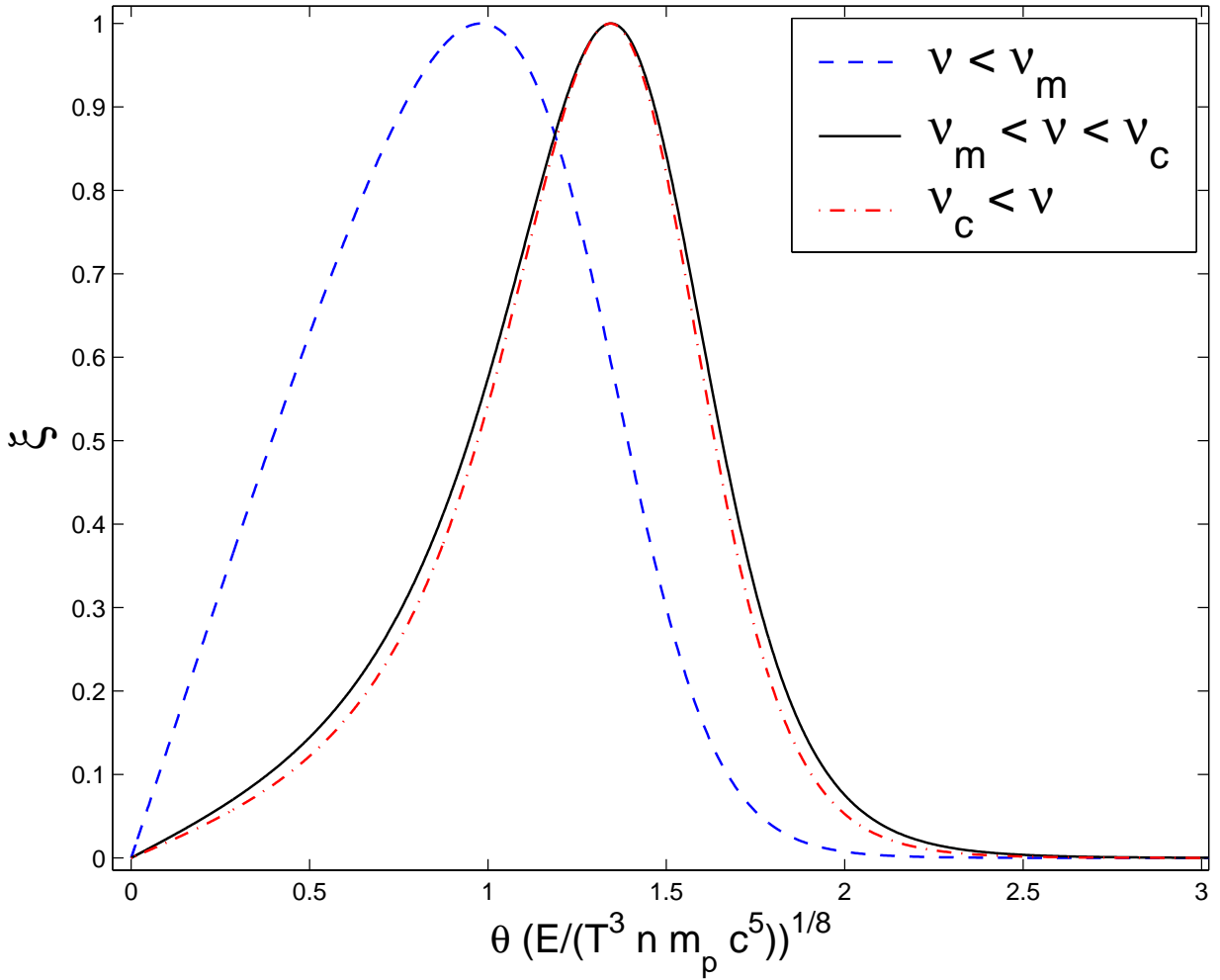}
\caption{The function $\xi(\theta)$ for the three spectral power
law segments. In this figure $p=2.2$, but $\xi$ is almost
insensitive to $p$ in the range $2<p<3$.}
\label{plotone}
\end{figure}

\section{GRB 021004}
The AG of GRB 021004 was observed on October 4'th 2002 at a
redshift of 2.32. The early optical detection (Fox et al. 2002),
$T \sim 0.005 \rm days$, enabled a detailed observation of this
afterglow from a very early stage. This unusual afterglow shows
clear deviations from a smooth temporal power law decay. A first
bump is observed at $T \sim 0.05 \rm days$, this bump is followed
by a very steep decay. Another smaller bump is observed at $T \sim
0.8 \rm days$ and a possible third one at $T \sim 3 \rm days $. A
steepening that may be a jet break is observed at $T \sim
4-7 \rm days$. During the first two days the optical spectrum is rather
constant (Pandey et al. 2003). Later, during the third bump and
the start of the break, the AG shows color variations (Matheson et
al 2003, Bersier et al 2003). This peculiar AG shows rapid
polarization fluctuations as well (both in degree and angle).
Between 0.3-0.8 days the polarization shows a fast drop and rise
combined with a rotation of $60^o$ (Rol et al 2003, Covino et al.
2002a; Wang et al. 2003). These fluctuations are correlated to the
light curve's fluctuations: at 0.3 days the light curve is at the
steep decay after the first bump while after 0.6 days it is at the
rise of the second bump. Another measurement after $\sim 4\rm
days$ shows another drop in the polarization level and a rotation
of $30^o$ (Covino et al. 2002b). While the last measurement
is taken at the beginning of the jet break and might be the result
of a jet seen off axis (Gruzinov 1999; Ghisellini \& Lazzati 1999;
Sari 1999; Rossi et al. 2002), the earlier measurements are taken
long before the jet break time and cannot be attributed to any of
these models. These models are unable to explain the observed rotation
(Lazzati et al 2003). The existence of 
rapidly varying polarization at such early stages indicates that
the axisymmetry of the flow is broken in a non-regular manner on
small angular scales. Here (fig 2) we show that the patchy shell
model can produce a variable light curve and polarization
and especially the angle rotation.

Several different mechanisms were suggested to explain this light
curve (Lazzati et al. 2002; Nakar et al. 2003; Holland et al.
2003; Pandey et al. 2002, Bersier et al. 2003; Schaefer et al.
2003; Heyl \& Perna 2003; Li \& Chevalier 2003,Kobayashi, S. \& Zhang
2002). Nakar \& Piran 
(2003) have shown that as a result of angular effects none of the
suggested spherical symmetric mechanisms can produce the steep
decay ($\sim t^{-1.5}$) observed after the first bump. This implied lack
of spherical symmetry  is strongly supported by the polarization
observations. Here we consider a symmetry break by a patchy shell.
Within this model the most natural
magnetic field configuration that produces correlated fluctuations
in the light curve and the polarization is the random field (see \S 2).

We have applied the solution presented in Eqs. \ref{EQ dFnu2} \&
\ref{EQ gamma} in a search for a reasonable angular energy
distribution that simultaneously produces the observed optical
(R-band) light curve and polarization. We expect such a
distribution to have a single characteristic angular scale
$\theta_{fl}=\pi/k_{max}$ and a contrast on the order of a few, as
was argued above. A random set of components was selected in two
dimensional Fourier space, with a cutoff at $k_{max}$ and a power
law spectral envelope $k^s$. The logarithmic contrast $c$ was defined such
that $r.m.s.(log_c(E/E_0))=1$, where $E_0$ is the typical energy.
We compare our results to the observed light curve during the
first two days. We assume that during this time the optical band
is between $\nu_m$ and $\nu_c$. The color changes during the third
bump and the following jet break prevent us from applying our
solution to later times.

\begin{figure}[h]
\plotone{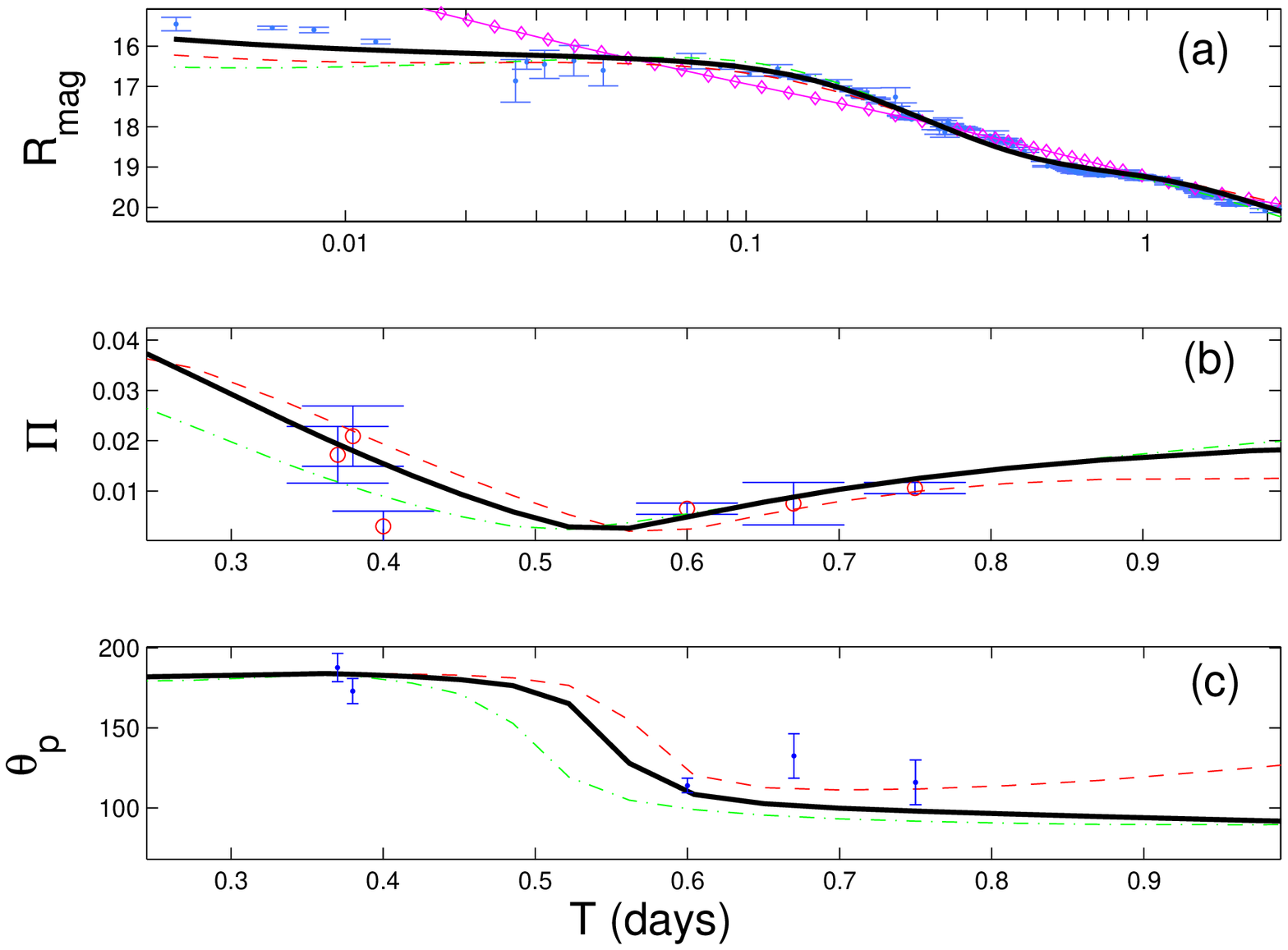}
\caption{The light curve (upper panel), polarization level (middle
panel) and angle (lower panel) obtained from three different random
energy distributions vs. the observations  of GRB 021004 ( light
curve: Fox et al. 2003, Uemura et al. 2003, Pandey et al. 2003,
Holland et al. 2003, Bersier et al. 2003; polarization (after ISM
correction: Rol et al 2003, Covino et al. 2002a; Wang et al.
2003). The straight line (marked with diamonds) in the upper panel
is a light curve obtained from an energy distribution randomly
generated using the same parameters. The same AG may appear either 
as a smooth or fluctuating power-law to different observers.}
\label{plottwo}
\end{figure}

Our strategy in trying to find a match between the model and the
observed light curve was to scan the $\{k_{max},c,s\}$ parameter
space and try to find the most suitable set of parameters.
According to $X$ and $\gamma$ ray observations we have used
throughout $p=2.2$ and $E_0=6 \cdot 10^{52}$ergs. For each such
set we produced $\sim100$ synthetic light curves. Each point in
this parameter space produces light curves with characteristic
time and amplitude scales. An agreement to the scales observed in
GRB 021004 was apparent in a relatively small neighborhood of
parameters, namely $\theta_{fl} \approx 0.017\rm rad$ (the wave
length of the fluctuations is $\approx 0.035\rm rad$ and $k_{max}
\approx 185 \rm rad^{-1}$), sharp spectrum, $s>2$, and a contrast
of $2.5<c<5$. These results are similar to the one obtained by
Nakar et al. 2003 with a much simpler model. We then visually
selected from the light curves in this neighborhood (covering
$\sim$ 1000 simulated lightcurves) three of those best fitting the
observed light curve, which are displayed in fig. 2. Very
reassuringly, those three energy profiles produce also a good fit
to the observed polarization (see fig. 2b). In agreement with the
observations, the polarization angle rotates by $45^o-80^o$
between 0.3-0.7 days (fig 2c). When fitting the polarization $b$,
the anisotropy parameter, is a free parameter. We find that in
order obtain the observed level of polarization $b\approx 0.5-0.8
[1.25-2]$ if the magnetic field is mainly planar [parallel]. This
$b$ decreases the level of polarization by a factor of $3-7$
compared to the maximal polarization obtained with $b= \infty$,
and this result is consistent with the low observed value of
polarization usually seen near the time of the jet break ($<3\%$)
compared to the expected value of $10-20\%$ (Sari 1999, Ghisellini
\& Lazzati 1999). The obtained value of $k_{max}$ justifies our
``frozen shell'' approximation. After two observer days $\Gamma
\approx 10$, hence $\theta_{fl} 
\gtrsim \theta_s \approx 1/40$ at all time ($T <2 \rm days)$.

\section{Conclusion}
Of the various models suggested to deal with fluctuations in GRB
AGs, we have dealt here with the "patchy shell" model. The
variability in this model results from the angular inhomogeneity
of energy in a shock-wave expanding into the circum-burst medium.
The time scale of these fluctuations is constrained to grow
linearly with time, namely $\Delta T \sim T$,
regardless of the angular scale of energy fluctuations in the
shell. There is also an amplitude decay, inherent in the smoothing
effect, which $\propto T^{-3/8}$. Another feature of this model is
a variable degree and direction of polarization resulting from the
azimuthal variation of the energy. The degree of polarization can
reach an order of tens of percents in the case of very anisotropic
magnetic fields.

As time progresses in the observer frame, radiation arrives from
larger $\theta$'s. Changes in the flux and polarization occur
when a group of fluctuations with a certain averaged orientation is
replaced by a new group with a different averaged orientation because
of this
change in the observed region. Therefore the transition from one peak
to the next in the light curve will characteristically be accompanied
by a rotation of polarization, with a drop in
polarization degree when the two groups contribute equally to the
flux. This drop will be less pronounced the closer the polarization
angle before and after the transition
is. Thus, the polarization variations are correlated
to the flux variations and occur on similar time scales. Note,
however, that a large rotation can take place on much shorter time scales.

The light curve and polarization of GRB 021004 are in agreement
with these general properties. Furthermore, we calculated a number
of light and polarization curves from a set of randomly generated
energy profiles and found recurring agreement between some of them
and the observed data. This model, however, fails to explain the
very short ($\sim1h$) time scale variations that might have been
observed at $T \sim 1 \rm day$ (Bersier et al. 2003), at least as long as
the frozen shell approximation holds, and there are no radial
variations in the energy.

An important prediction arising from the self similar flux profile
is a logarithmic time lag between light curve and polarization
variations below and above $\nu_m$. A more accurate analysis of
this problem, which we are currently carrying out, can be made by
taking into account the finite thickness and the hydrodynamic
profile of the radiating area and performing a three dimensional
integration of the flux originating from different radii.

We would like to thank Re'em Sari and Davide Lazzati for helpful
discussions. We especially thank Tsvi Piran for insightful
remarks. The research of EN was partially supported by the Horowitz
foundation and by the generosity of the Dan David prize by the Dan
David scholarship 2003.

\end{document}